# Interactive visualization of kidney micro-compartmental segmentations and associated pathomics on whole slide images

Mark S. Keller[1], Nicholas Lucarelli[2], Yijiang Chen[3,4], Samuel Border[2], Andrew Janowczyk[5], Jonathan Himmelfarb[6,7], Matthias Kretzler[8], Jeffrey Hodgin[8], Laura Barisoni[9,10], Dawit Demeke[11], Leal Herlitz[12], Gilbert Moeckel[13], Avi Z. Rosenberg[14], Yanli Ding[15], for the Kidney Precision Medicine Project and HuBMAP Consortium, Pinaki Sarder*,[2], *Nils Gehlenborg*[,1]

[1] Department of Biomedical Informatics, Harvard Medical School, Boston, MA, USA
[2] Department of Medicine - Quantitative Health Section, University of Florida, Gainesville, Florida, USA
[3] Center for Computational Imaging and Personalized Diagnostics, Case Western Reserve University, Cleveland, OH, USA
[4] Department of Radiation Oncology, Stanford University, Stanford, CA, USA
[5] Department of Biomedical Engineering, Emory University and Georgia Institute of Technology, Atlanta, GA, USA
[6]Department of Medicine, Icahn School of Medicine at Mount Sinai, New York, NY, USA
[7]Center for Kidney Disease Innovation at Icahn School of Medicine at Mount Sinai, New York, NY, USA
[8]Department of Medicine, University of Michigan, Ann Arbor, MI, USA
[9] Department of Pathology - Division of AI and Computational Pathology, Duke University, Durham, NC, USA
[10] Department of Medicine - Division of Nephrology, Duke University, Durham, NC, USA
[11]Department of Pathology, University of Michigan, Ann Arbor, MI, USA
[12]Department of Anatomic Pathology, Cleveland Clinic, Cleveland, OH, USA
[13]Department of Pathology, Yale University School of Medicine, New Haven, CT, USA
[14]Department of Pathology, Johns Hopkins University School of Medicine, Baltimore, MD, USA
[15]Department of Laboratory Medicine and Pathology, University of Minnesota, Minneapolis, MN, USA

* indicates equal contributions, and corresponding authors: pinaki.sarder@ufl.edu & nils@hms.harvard.edu

# Abstract

Application of machine learning techniques enables segmentation of functional tissue units in histology whole-slide images (WSIs). We built a pipeline to apply previously validated segmentation models of kidney structures and extract quantitative features from these structures. Such quantitative analysis also requires qualitative inspection of results for quality control, exploration, and communication. We extend the Vitessce web-based visualization tool to enable visualization of segmentations of multiple types of functional tissue units, such as, glomeruli, tubules, arteries/arterioles in the kidney. Moreover, we propose a standard representation for files containing multiple segmentation bitmasks, which we define polymorphically, such that existing formats including OME-TIFF, OME-NGFF, AnnData, MuData, and SpatialData can be used. We demonstrate that these methods enable researchers and the broader public to interactively explore datasets containing multiple segmented entities and associated features, including for exploration of renal morphometry of biopsies from the Kidney Precision Medicine Project (KPMP) and the Human Biomolecular Atlas Program (HuBMAP).

# Introduction

Application of machine learning and artificial intelligence to bioimaging is transforming pathology and making computational pathology a reality. Advancements in computational image analysis now make it possible to perform panoptic segmentation in which instance-type and semantic-type segmentations can be computed simultaneously. Current state-of-the-art methods can achieve comparable accuracy to pathologists enabling use of the resulting data for clinical assistance. Confidence in segmentation results has opened the door to derivation of instance-level features, measurements termed pathomics. The datasets that result from application of these segmentation and pathomic feature extraction methods are complex and consist of histology images, multiple types of segmentation bitmasks (i.e., label images), pathomic observation-by-feature matrices, and categorical or hierarchical labels for observations. Further complexity of datasets may come in the form of network/graph-based representations or machine learning-based predictions of other 'omics measurements, such as transcriptomics, proteomics, metabolomics, to name a few.

Consortia such as the Kidney Precision Medicine Project (KPMP)[1] are applying computational pathology methods to patient biopsies on a large scale to study how diseases, such as chronic kidney disease (CKD) and acute kidney injury (AKI) affect renal pathology. Researchers aim to combine computational pathology with statistical methods and complementary multimodal omics experiments to improve diagnosis, prognosis, and treatment for kidney patients. Similar efforts are underway for other diseases such as in the Human Tumor Atlas Network (HTAN)[2] and for normal reference tissue in the Human Biomolecular Atlas Program (HuBMAP)[3].

KPMP is making hundreds of computational pathology datasets available to researchers, patients and the broader public. This wide audience has the common goal of developing a deeper understanding of kidney disease and reference kidney function using the rich and high-quality data generated by the consortium. However, the individuals in this audience have differing levels of computational expertise. Many may lack familiarity with the specialized software tools and file formats commonly used in bioinformatics and computational pathology. This gap presents obstacles for non-computational researchers, patients, and others who aim to make sense of the data.

Even for researchers with computational skills, the scale, complexity, size, and number of computational pathology datasets can pose obstacles to their analysis. In many cases it may not be feasible or practical for researchers to download data locally due to storage constraints and/or privacy concerns. Researchers may need to select a subset of biosamples of interest prior to downloading data for further analysis. To minimize time spent on data cleaning and data conversion, computational pathology results should be shared using open standard file formats that are efficient (in terms of storage and access) and compatible with a wide range of software tools. Infrastructure costs and maintenance time should also be minimized so that researchers can focus on extracting biologically- and clinically- relevant insights.

The rise in complexity and size of datasets in computational pathology has not been matched by establishment of standard data representations or the development of interactive visualization tools. In contrast, standards and tools which support either bioimaging data or single-cell omics data alone are actively being developed by researchers in biology and computer science [4,5]. Visualization tools for bioimaging data often lack support for categorical and quantitative measurements associated with segmented entities. On the other hand, visualization tools for spatially-resolved -omics data often support cellular entities and lack support for both **multiple observation types** (e.g., cells, tubules, glomeruli) and **multiple data modalities** (e.g., transcriptomics and pathomics). Many existing tools in both domains (i.e., bioimaging and single-cell -omics) require specialized servers, non-standard data formats, or are implemented in the form of desktop applications that must be installed.

To overcome these obstacles, we developed a web-based interactive visualization tool that loads computational pathology data from the cloud. In conjunction, we propose a standard representation for segmentations of multiple types of observations, which can be linked to pathomics or other -omics measurements at the observation level (e.g., expression level of a cell, or aspect ratio of a peritubular capillary). We applied this approach to KPMP kidney biopsies processed through a previously described deep learning-based segmentation and pathomic feature extraction pipeline[6,7]. We show that this interactive visualization approach makes computational pathology pipeline outputs for hundreds of samples accessible to a broad audience of data consumers (i.e., accessible directly within web-based data portals, without the need to download or install specialized software).

# Results

**Interactive visualization of segmentations and associated pathomics features.** To support visualization of computational pathology data, we have extended Vitessce, a web-based interactive visualization framework for -omics and bioimaging data [8].

The Vitessce framework was originally designed for spatially-resolved single-cell -omics data visualization. Computational pathology datasets share many characteristics with single-cell datasets but differ in the scale and number of types of entities. For example, spatial single-cell experimental methods such as multiplexed immunofluorescence result in datasets containing high-resolution images, segmentation bitmasks, and cell-by-channel expression matrices. These data types are analogous to the whole-slide images, segmentation bitmasks, and observation-by-feature matrices in pathomics. However, many tools designed for single-cell data consider and support only one type of observation – cells – and one type of feature – genes. In contrast, pathologists must consider information from different functional tissue units, with varying organizational and cellular composition complexity.

Vitessce supports simultaneous visualization of datasets containing multiple types of observations and features. Segmentations of functional tissue units can be layered on top of one another, with different types of segmented entities assigned to different visual properties. For example, segmentations of tubules (one observation type) can be assigned colors based on a "tubular luminal fraction" feature variable, while glomeruli (a different observation type) can be colored based on whether they are classified as "globally sclerotic" versus "non-globally sclerotic".

Other characteristics of the Vitessce framework inform our choice to use it for the computational pathology visualization use case. As web-based software, it does not require specialized server infrastructure or installation of a desktop application. Large images and other data files can be loaded from cloud object storage systems or standard web file servers. Vitessce can be used not only in web applications but as an interactive widget in computational notebooks such as JupyterLab and RMarkdown documents.

**Establishment of standard representation for segmentations of multiple observation types.** Along with the visualization software, we propose a standard way to represent and organize segmentation data which comprises multiple types of observations (Figure 1b). Rather than developing a specialized file format, we build upon established formats for tabular and imaging data.

We propose the following approach, which simply takes the form of a convention that leverages existing open standard formats:
- **Images**: Store RGB or multiplexed imaging data using open standards: OME-TIFF [9,10] or OME-NGFF [4,11] (including OME-NGFF images contained within SpatialData [5] objects).
- **Segmentations**: Store segmentation bitmasks in OME-TIFF[9,10] or OME-NGFF[4,11] format (including within SpatialData Labels elements) in which each channel corresponds to one observation type (Figure 2a). Pixel values start from 1 and correspond to instances of observations, with 0-value pixels representing background. Channel names should specify which channels correspond to which observation types (e.g., cell, tubule, glomerulus). Alternatively, segmentations can be stored as polygons using SpatialData Shapes elements.
- **Pathomics**: Store pathomic features in open standard tabular formats: CSV, AnnData [12,13] (including within SpatialData Tables elements), or MuData[14]. Each table should correspond to a single observation type (i.e., bitmask channel). There must be an additional data structure to map between tables and observation types – this is included in the MuData or SpatialData formats (which are containers for multiple tables) but may require an additional file in the case of single-table formats such as CSV and AnnData.

The OME-TIFF and OME-NGFF formats are familiar to researchers performing and analyzing microscopy data. CSV is a widely-used tabular format. AnnData & MuData are efficient tabular formats that are commonly used by single-cell biologists but are designed more generally to represent observations that are accompanied by rich metadata such as feature vectors and categorical class labels. As the SpatialData[5] format becomes adopted within the bioimaging and -omics ecosystems, we anticipate that SpatialData will also become useful for storage of computational pathology data.

The abstract nature of our proposed data representation approach makes it possible to be used with both current and future file formats. While we provide a reference implementation of a visualization tool that supports this representation, we anticipate that other tools and methods at the intersection of computational pathology and spatial -omics will adopt this format.

**FTU segmentation and associated pathomics features.** We demonstrate our visualization and data representation approaches by using them to visualize the results of a state-of-the-art pipeline for segmentation and feature extraction of functional tissue units (FTUs) in the human kidney (Figure 1a). The development, training, and performance characteristics of this deep-learning based segmentation approach were previously published [15].

This pipeline supports segmentation of multiple microcompartments within the kidney, including cortical and medullary interstitium, non globally sclerotic glomeruli, globally sclerotic glomeruli, tubules, arteries/arterioles [7,15], and peritubular capillaries (PTC) [6]. Regions of interstitial fibrosis and tubular atrophy (IFTA), generated through manual segmentation by KPMP study pathologists, can also be visualized alongside the repertoire of automatically segmented FTUs.

From the segmented FTUs, the pipeline extracts associated pathomic features, including morphological, texture, color and distance transform features. Feature extraction is performed using a traditional image analysis approach (e.g., color thresholding) for interpretability [16].

**Application of approach and deployment**. We have applied the data storage and interactive visualization approach presented here both to biopsies individually and in an automated fashion across many biopsies from the KPMP and HuBMAP consortia via integration into centralized data processing and dissemination infrastructure.

As of January 8, 2024, we have applied the pipeline to 222 (PAS-stained) histology images from 112 KPMP biopsies, all of which are available to explore interactively via the Spatial Viewer within the KPMP Atlas data portal (Figure 2b). Of these 112 samples, 11 fit the clinical criteria (based on primary adjudicated category) for Acute Interstitial Nephritis, 16 for Acute Tubular Injury, 39 for Diabetic Kidney Disease, and 24 for Hypertensive Kidney Disease (the categories for the remaining samples were "Other" or "Cannot be determined"). This broad deployment of the approach showcases the ability of this pipeline to extract segmentations from both reference and disease samples at the atlas scale.

In the form of a case study, we next demonstrate how characteristics of an individual biopsy can be identified and analyzed via visualization in Vitessce of the FTU segmentation and associated pathomic features (Figure 2c). This process involves the following steps:

(1) Visualization of the whole slide image for a kidney biopsy of interest. This can be achieved by navigating to Vitessce within a data portal or by configuring Vitessce via its online editor or via code in the JavaScript, Python, or R programming languages.
(2) Visualization of the segmentations for the PTCs as a layer on top of the image. This can be achieved by toggling their visibility using the layer controller. The zoom in function allows loading higher resolution data to view the segmentations in more detail. By default, the PTC segmentations are all mapped to a single color, which can be selected or changed by the user.
(3) Visualization of pathomic features mapped on segmented PTCs. Individual pathomic features can be mapped onto the segmented PTCs by selecting them from the list of columns available within the pathomic feature table. A quantitative color map can also be used to map pathomic features values in each segmented PTC. For example, we can select "area" from the list of features for PTCs, and we can reference the quantitative colormap legend to distinguish capillaries with larger area (in yellow) from those with smaller area (in dark purple).
(4) Spatial relationship between quantitative PTC pathomic features and the microenvironment. We can view regions of IFTA by toggling the visibility of this segmentation channel. We can adjust the transparency levels of the PTC and IFTA segmentation channels to view the image and all segmentations simultaneously.
(5) Hypothesis testing and analysis. Upon viewing the IFTA regions and PTCs with color-mapped area values, we hypothesize that the distribution of PTC area differs within IFTA regions versus non-IFTA regions. We can use linked interactive statistical plots to quickly compare these distributions, stratifying PTC quantitative characteristics (i.e., size) based on whether they lie within cortical IFTA, cortical non-IFTA, or total cortex regions.

## Discussion

Application of artificial intelligence and machine learning in pathology has brought about revolutionary advances and the emergence of the field of computational pathology [17,18]. Meanwhile, spatial -omics approaches are advancing in parallel. As it is of increasing interest to perform both histology and spatial -omics assays on identical or serial tissue sections, it is becoming increasingly important to be able to analyze these data modalities in tandem during the downstream data interpretation stages.

Many existing tools support interactive visualization for digital pathology and for -omics data – but few tools support visualization of these data types together. Related work in pathology visualization includes HistomicsUI, which supports many interactive features such as annotation, but requires specialized server-side infrastructure [19]. Slim supports viewing images in DICOM format via DICOMweb server-side infrastructure [20]. Tools such as Digital Slide Archive and OMERO help users to organize large collections of images with metadata [10,21]. The Comparative Pathology Workbench uses the "small multiples" visualization technique to show many images simultaneously [22].

Several visualization tools are designed for both pathology and multiplexed imaging data. For desktop-based tools, these include CellProfiler, Napari, ImageJ, and QuPath, which help users to perform image analysis using a combination of programmatic APIs and graphical user interfaces [23–25]. On the web, Minerva provides guided exploration of multiplexed images alongside narrative content and author annotations [26]. A number of image and cell-based analysis methods have been designed for close integration with these visualization tools, including MCMICRO for Minerva and SquidPy for Napari [27,28].

On the other hand, there are many related tools designed for visualization of spatially-resolved -omics data. This includes TissUUmaps [29], a scalable web-based tool for interactive visualization of single-cell and single-molecule 'omics data. Notably, in many such tools, cells and transcripts are the primary supported entities, with all current tools lacking support for visualization of other types of segmented observations (i.e., larger functional tissue units), and by extension they lack support for associating such segmentations with multiple observation-by-feature tables containing associated quantitative or qualitative measurements or derived features. There is a need for tools which support visualization of both computational pathology and spatial 'omics data types.

To enable the development of such tools (including ours), we contribute a new standard for storage of segmentation and pathomic feature data. We propose the reuse of existing file formats because there is much prior work in development of formats for imaging and 'omics data. OME-TIFF [10] and OME-NGFF [4,11] are two file formats that support multiplex multiscale images with rich experimental metadata. These formats also support storage of images that represent segmentation bitmasks. In the single-cell analysis ecosystem there are many formats designed around tables of observations and features that are accompanied by metadata, such as AnnData [30], MuData [14], SingleCellExperiment, and Seurat [31]. More recently, formats have emerged that are designed to contain both spatially-resolved data such as images and segmentations as well as observation-by-feature tables and their metadata (but not necessarily more than one type of observation) [5]. At the time of writing, the ecosystem of data analysis methods is fragmented in terms of supported file formats which motivates our higher-level approach that supports a combination of multiple underlying open-standard file formats. For example, CSV and AnnData can be interchanged for tabular data while OME-TIFF and OME-NGFF can be swapped for imaging and segmentation data, so long as there is a correspondence between data tables and types of segmented entities.

We demonstrate our approach on KPMP biopsies segmented by an exemplary pipeline [6,7]. However, the visualization approach that we propose is generic and could be reused with any segmentation pipeline that outputs results in the open standard formats as described in the Results section. Further, scripts could be developed to convert alternative data representations to those that we propose, or plugins can be developed for visualization tools to add compatibility for loading data from additional formats.

The Vitessce framework is robust and production-ready as it is already being used in data portals for large atlas-building consortia such as HuBMAP and KPMP to visualize data from spatial single-cell experiments. The framework is extensible via a plugin system that enables individuals with web development expertise to define custom views or support additional file formats. Vitessce uses the Viv library for scalable multiplex image visualization without a server [32].

This work demonstrates the importance of considering that tissues are composed of multiple functional tissue units when developing computational tools and data structures, despite the focus to date on cells (and subcellular components) and the established ecosystem of single-cell analysis tools. As deep learning-based segmentation methods advance for additional types of functional tissue units (FTUs) within other organs, visualization of the resulting FTU segmentations will become more valuable for purposes including exploration and communication. Network-based analyses will become increasingly important and therefore overlaid network visualizations are an important future direction for this work. Other follow-up work will be needed to support tasks such as proofreading (i.e., detection and correction of segmentation errors), human-in-the-loop annotation, and support of additional dimensions (e.g., 3D, temporal) and scales.

# Figures

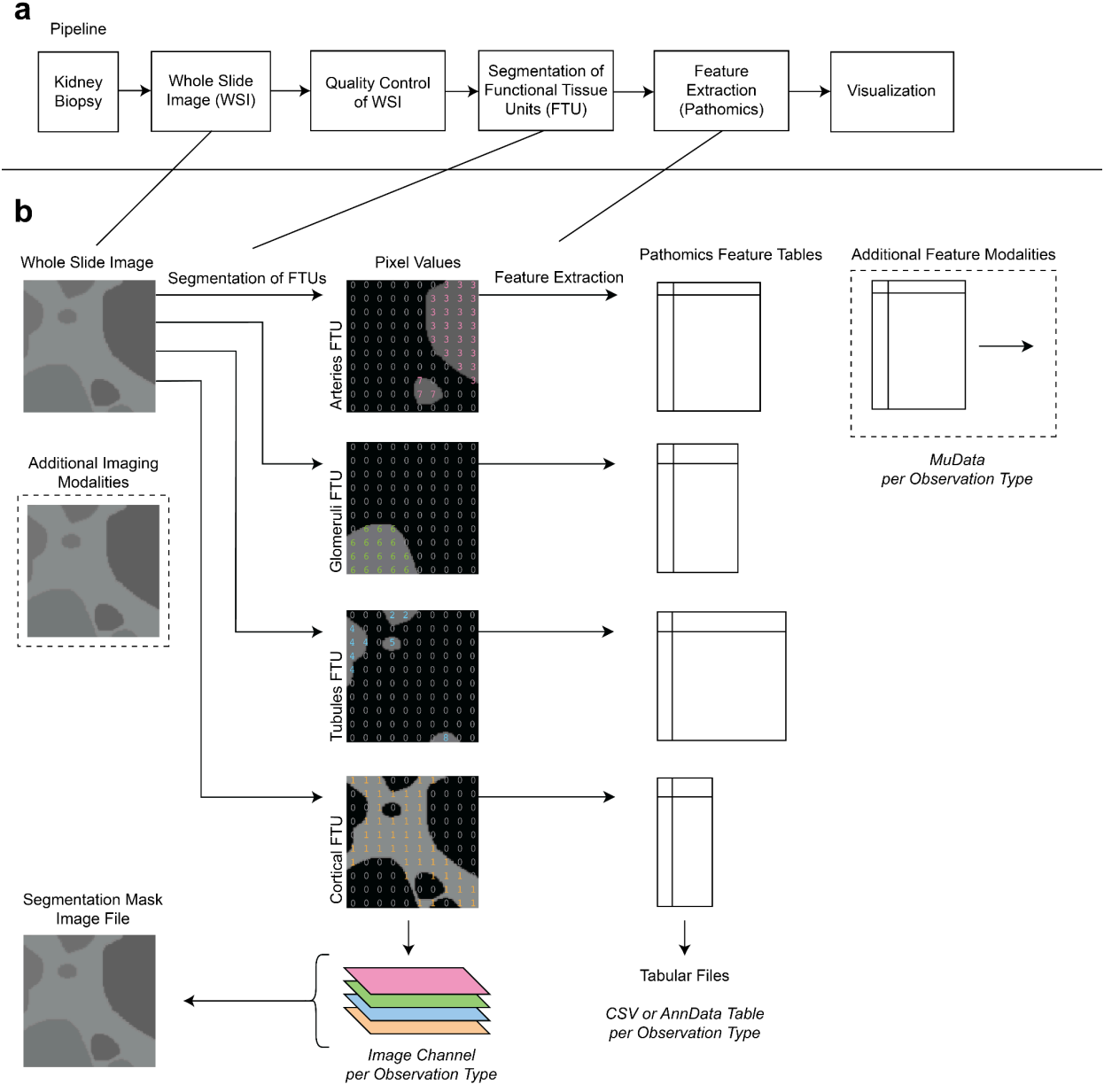

Figure 1. a) Overview of pipeline from biopsy and whole slide image (WSI) to segmentation and pathomics to visualization with Vitessce. b) Segmentation masks are represented as multi-channel image files with one channel for each type of segmented entity ("observation type"). For instance, the first channel contains segmented arteries (observation type: artery) while the second channel contains segmented glomeruli (observation type: glomerulus). Pathomics and other features are represented with one table per feature and observation type. Feature tables contain one row per observation (e.g., in the artery feature table, the first row corresponds to the first segmented artery; in the glomerulus feature table, the first row corresponds to the first segmented glomerulus) such that different feature tables contain different numbers of rows. Feature table row indices are expected to match the pixel values

within the corresponding segmentation mask channel (e.g, the artery corresponding to arteries feature table row index 1 has pixel value 1 in the segmentation mask image channel for arteries).

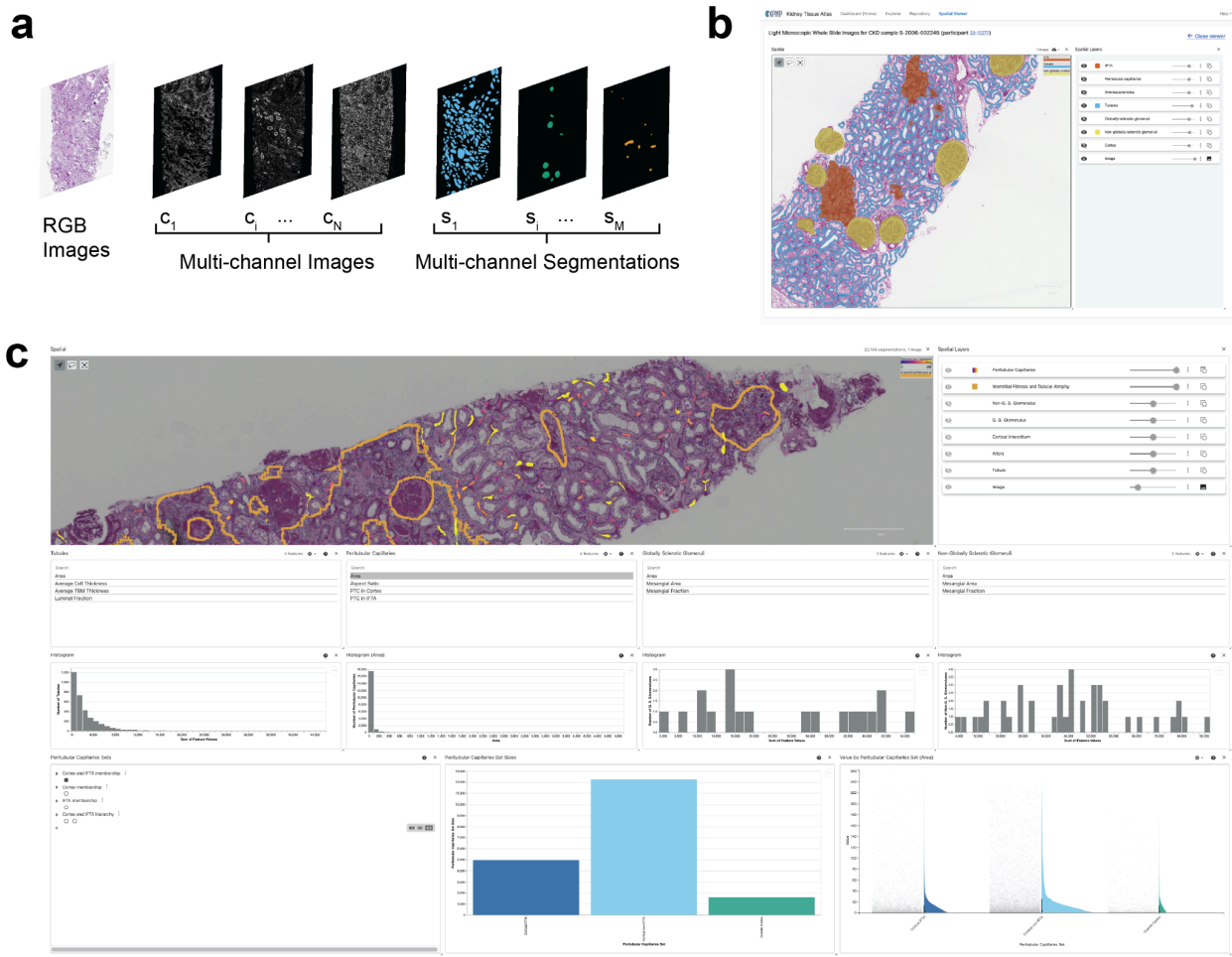

Figure 2. a) Schematic of superimposed approach to visualization of image and segmentation layers. b) Screenshot of visualization interface embedded in the KPMP Kidney Tissue Atlas web portal, displaying segmentation masks for a sample of interest. c) Screenshot of Vitessce visualization interface during exploration of a single biopsy with segmentations of functional tissue units (FTU) displayed and extracted features visible in both histograms and color mappings.

URL: https://vitessce.io/#?dataset=sdata-kpmp-2023

Alternate URL: https://legacy.vitessce.io/demos/2025-08-21/9ef3f70f/?dataset=sdata-kpmp-2023

# Methods

## Extension of Vitessce

Vitessce as described by Keller et al. [8] supports spatially-resolved visualizations (e.g., segmentations) comprising a single type of observation (e.g., cells). We extended the spatial view in Vitessce, generalizing it to support multiple observation types (e.g., multiple types of functional tissue units) simultaneously.

### Visualization rendering

We implemented a WebGL fragment shader (compatible with the DeckGL framework used by Vitessce) that renders a multi-channel bitmask image. This shader supports dynamic color mapping on a per-channel level, for both categorical and quantitative colormaps. For example, a user may want to color the Glomeruli bitmask in image channel 0 using a categorical colormap (sclerotic and non-sclerotic categories) while coloring the Tubules bitmask in channel 1 using a quantitative colormap (continuum of luminal fraction values). The shader supports specification of the following visual properties on a per-channel basis: "isFilled", "opacity", "strokeWidth", "colormapRangeStart", "colormapRangeEnd", "isStaticColorMode", "isSetColorMode", "isVisible". In staticColorMode, a channel can be mapped to its own RGB color value which will be used to render all non-zero (i.e., non-background) pixels. In setColorMode, a channel can be passed a mapping between sets of observations and RGB color values (i.e., one color per set ID). In featureColorMode, the shader applies a quantitative colormap function (e.g., magma or viridis) to each observation using a feature value vector (i.e., one feature value per observation) for the channel. Each channel can be assigned to a different color mode; the color mode is specified per channel. Provided RGB values, set membership values, and feature values are channel-specific. In order to pass these values to the GPU, they are stored efficiently using WebGL textures.

### Algorithm for on-the-fly stroked vs. filled bitmask rendering

Bitmask images associate pixels with observation IDs using a "filled" approach (i.e., all pixels corresponding to a cell are assigned the cell ID value). However, it is often desired to render these using a "stroked" approach in which only the observation outline is colored while the interior remains transparent.

A naive approach is to pre-determine the pixels along each cell's outline, and construct a separate bitmask image in which the interior pixels of each cell are represented using the background pixel value. While this approach will result in the interior pixels being mapped to a transparent color and therefore only rendering the cell outlines, it has several drawbacks. The application must load two separate bitmask images, one for filled and one for stroked visualization, for each bitmask. These two images must be precomputed and stored, requiring double the disk space and complicating visualization configuration. Finally, the width of the cell outline must be fixed at the time of bitmask construction. For these reasons, we devised an on-the-fly stroked rendering method for filled bitmask images.

This algorithm is implemented as a function within a WebGL fragment shader and is therefore executed for every pixel via the GPU. The function takes as input a single channel bitmask stored as a WebGL texture, the current pixel coordinate, and the user-defined stroke width. Then lookups must be performed for the current pixel value ("currentValue") as well as the values of the pixels located strokeWidth-pixels away in eight different directions (north, south, west, east, north-west, north-east, south-west, south-east). The function then compares currentValue to the eight other pixel values. If one or more of the eight values differ from currentValue, then we conclude that the current pixel is located less than strokeWidth-pixels away from the edge of the segmentation boundary and therefore should be rendered opaque as opposed to transparent (during stroked rendering mode). Within the same fragment shader, we execute this function for each channel of multi-channel segmentation bitmasks and combine the results with user-defined per-channel properties such as colormaps and opacity values.

### Configuration of spatial and layer controller views

We implemented spatial and layer controller views for Vitessce which support a hierarchy of coordinated properties. For example, each segmentation bitmask layer has properties such as "isVisible," with each layer having multiple children segmentation channels (corresponding to functional tissue units) that each have their own visual properties such as "color" and "isFilled." This hierarchical approach to visualization configuration and coordination follows the recent work described by Keller *et al.* [33].

### Data loading

To facilitate configuration of multiple segmentation bitmasks contained in the same image file (e.g., multiple channels of an OME-TIFF), we use channel name metadata fields to specify observation types (i.e., the type of functional tissue unit). For example, a two-channel segmentation bitmask image may have the channel names "glomerulus" and "tubule." Then, Vitessce uses the channel name metadata from the image to determine which channels correspond to each observation type and identify correspondences between channels and other files (e.g., tables containing extracted feature values for glomeruli and tubules).

## Development of models for segmentation

Three previously validated deep learning-based segmentation models were used in the segmentation of renal structures and pathology-associated regions in the kidney. Each model was trained on manual and bootstrapped annotations validated by a renal pathologist.
The first model is outlined in Lucarelli *et al.* [15], and supports segmentation of both the cortical and medullary interstitia, non-globally-sclerotic glomeruli, globally-sclerotic glomeruli, tubules, and arteries/arterioles. This model is built on the Detectron2 panoptic segmentation architecture. Kidney samples from various institutions, various stain types, and various disease states, including diabetic nephropathy, lupus nephritis, transplant surveillance, and reference tissues were included in the training of the network. This model was previously validated in a PAS-stained holdout set.

The second model is outlined in Jayapandian *et al.* [6], and supports segmentation of the peritubular capillaries. This network is built on the standard U-Net architecture, and was trained using a large sample of the NEPTUNE cohort that contained four different histological stains from patients with Minimal Change Disease (MCD). This model was previously validated in a holdout set with each of the same four stains.

The last model is outlined in Ginley *et al.* [7], and supports segmentation of the pathology-associated regions of Interstitial Fibrosis and Tubular Atrophy (IFTA). This model is built on the DeeplabV2 semantic segmentation architecture, and was trained using a large, multi-institutional cohort of PAS-stained renal biopsies from patients with Diabetic Nephropathy, transplant kidneys, deceased-donor kidneys, and reference kidneys. The model was previously tested in a PAS-stained holdout set, and compared to multiple pathologist annotations for agreement.

## Extraction of pathomics features

Hand-engineered pathomic features were designed with the help of renal pathologists for clinical relevance and explainability. These features can be classified as morphological features, texture features, color features, and distance transform features. Sizes, shapes, color means and standard deviations, and thicknesses of sub-segmentations make up the majority of the feature set.

## Specification for on-disk representation

Label images may be stored using OME-TIFF or OME-NGFF (including OME-NGFF images contained within the Labels element of a SpatialData object). Each channel represents a different type of observation. Channel names indicate the type of observation contained in the channel. The pixel value zero is used to indicate background. A segmentation for the ith observation of a given type is indicated by pixels with value i.

Each observation (e.g., instance of an FTU such as a glomerulus) may be associated with feature values (i.e., an observation-by-feature matrix for each channel, where each row corresponds to an observation belonging to the channel's observation type). These observation-by-feature matrices may be stored in an AnnData object, MuData object, or CSV file. The observation index values for each observation-by-feature matrix should match the non-zero pixel values within the corresponding label image (i.e., should consist of 1-indexed integers, although not necessarily be ordered sequentially or gapless).

## Acknowledgements

The Kidney Precision Medicine Project (KPMP) is supported by the National Institute of Diabetes and Digestive and Kidney Diseases (NIDDK) through the following grants: U01DK133081, U01DK133091, U01DK133092, U01DK133093, U01DK133095, U01DK133097, U01DK114866, U01DK114908, U01DK133090, U01DK133113, U01DK133766, U01DK133768, U01DK114907, U01DK114920, U01DK114923, U01DK114933, U24DK114886, UH3DK114926, UH3DK114861, UH3DK114915, and UH3DK114937. We gratefully acknowledge the essential contributions of our patient participants and the support of the American public through their tax dollars. This work was supported by additional funding from the National Institutes of Health (R01 DK114485, P.S.; R21 DK128668, P.S.; OT2 OD033753, P.S.; R01 DK129541, P.S.; R01 AR080668, P.S.; OT2OD026677, N.G.; OT2OD033758, N.G.; T15LM007092, N.G.; R33CA263666, N.G.; T32HG002295, M.S.K.). The content is solely the responsibility of the authors and does not necessarily represent the official views of the National Institutes of Health. The authors acknowledge the University of Michigan Medical School Central Biorepository (RRID:SCR_026845) for providing biospecimen storage, management, and distribution services in support of the research reported in this publication/grant application/presentation. The authors acknowledge Stephen Fisher for inspiring the diagram in Figure 1b.

## Competing interests

N.G. is a co-founder and equity owner of Datavisyn. A.J. provides consulting for Merck, Lunaphore, and Roche, the latter of which he also has a sponsored research agreement. P.S. is an advisor for DigPath Inc. The other authors declare no competing interests.

## Data availability

The data described in this manuscript is available from the KPMP Atlas data portal at https://atlas.kpmp.org/spatial-viewer/?filters[0][field]=imagetype&filters[0][values][0]=PAS (Segmentation Masks)&filters[0][type]=any.

## Code availability

The JavaScript and WebGL code which implements the interactive visualization features described here are contained in the Vitessce repository at https://github.com/vitessce/vitessce. Segmentation and feature extraction code are contained in the repositories at https://github.com/SarderLab/Multi-Compartment-Segmentation, https://github.com/SarderLab/IFTA_segmentation and https://github.com/SarderLab/CombinedFeatureExtraction. PTC segmentation and feature extraction code are available at https://github.com/yijiangchen/DL-kidneyhistologicprimitives and https://github.com/yijiangchen/PTC_feature_extraction.